\documentclass[12pt]{iopart}

\expandafter\let\csname equation*\endcsname\relax
\expandafter\let\csname endequation*\endcsname\relax

\usepackage{graphicx}
\usepackage{
amsfonts,
amssymb,
amsmath,
latexsym,
epsfig,
wasysym,
bbm
}

\usepackage{bm}
\newcommand{\ff}{\bm{f}}
\newcommand{\dd}{\bm{\Delta}}
\newcommand{\pp}{\bm{p}}
\newcommand{\qq}{\bm{q}}

\usepackage{color}
\definecolor{bluecolor}{rgb}{0,0.,1.}

\definecolor{redcolor}{rgb}{.7,0.,0.}

\begin{document}

\title{Generalized Entropies and the Similarity of Texts}

\author{Eduardo G. Altmann}
\address{School of Mathematics and Statistics, University of Sydney, 2006 NSW, Australia}
\address{Max Planck Institute for the Physics of Complex Systems, D-01187 Dresden, Germany}
\ead{ega@maths.usyd.edu.au}

\author{La\'{e}rcio Dias}
\address{Max Planck Institute for the Physics of Complex Systems, D-01187 Dresden, Germany}

\author{Martin Gerlach}
\address{Max Planck Institute for the Physics of Complex Systems, D-01187 Dresden, Germany}
\address{Department of Chemical and Biological Engineering, Northwestern University, Evanston, Illinois 60208, USA}

\begin{abstract}
We show how generalized Gibbs-Shannon entropies can provide new insights on the statistical properties of texts.  The universal distribution of word frequencies
(Zipf's law) implies that the generalized entropies, computed at the word level, are
dominated by words in a specific range of frequencies. Here we show that this is the case
not only for the generalized entropies but also for the generalized (Jensen-Shannon) divergences,
used to compute the similarity between different texts.  
This finding allows us to identify the contribution of specific words (and word frequencies) for the different generalized entropies and also to estimate the size of the databases needed to obtain a reliable estimation of the divergences. 
We test our results in large databases of books (from the Google n-gram database) and scientific papers (indexed by Web of Science).
\end{abstract}

\maketitle

\tableofcontents

\section{Introduction}

Generalized entropies, such as the Renyi and Tsallis entropies, have been studied in
different aspects of Statistical Physics~\cite{gellmann.book2004,BookByTsallis} and Non-linear Dynamics~\cite{kantz.book2003}. In Information Theory,
these entropies are viewed as a generalizations of the Shannon entropy that are potentially
useful in particular problems.  Many problems require the comparison of the divergence
(or, its opposite, the similarity) between two or more signals, a problem that can be
quantified through the use of divergence measures based on generalized (joint) entropies, e.g. in analysis of DNA sequences~\cite{grosse.2002} or image processing~\cite{he.2003}.

A traditional and increasingly important application of information theory is the analysis
of (signals based on) natural language~\cite{manning.book1999,boyack.2011,masucci.2011a,dodds.2011,bochkarev.2014,pechenick.2015}. This analysis often happens at the level of
words, i.e., in which each word (type) is considered a different symbol of analysis. One
important statistical feature in the statistical analysis of word frequencies is the
existence of linguistic laws~\cite{AltmannGerlach}, i.e., statistical regularities 
observed in a variety of databases.  The most famous case is Zipf's law, which specifies how the frequencies of words are distributed~\cite{zipf.book1936,gerlach.2013,piantadosi.2014,moreno.2015}. 

In this paper we explore the implications of linguistic laws to the computation of
information-theoretic measures in written text.
While information-theoretic approaches typically measure the similarity of an ensemble of words (the vocabulary), we show how generalized entropies can be used to assess the influence of individual words to these (global) measures, providing a bridge to the studies on evolution of language following trajectories of individual words~\cite{pagel.2007,lieberman.2007}.
In particular, we show how the contribution of individual words, appearing in different scales of frequency, vary in the
different generalized entropies. We explore the implications of our findings to two problems: (i) the best generalized entropy for highlighting the contribution of Physics keywords; and (ii) determining how large a given database has to be in order obtain sufficient coverage/sampling of the generalized entropies.

\section{Basic concepts}

We are interested in extracting information about written documents based on the number of times $N_i$ each word $i=1, \ldots, M$ appears in each database. For each database, we denote by $f_i$ the frequency of the word $i$ (i.e., $f_i \equiv N_i / \sum_{i=1}^M N_i$), which we consider to be an estimator of the probability $p_i$ of occurrence of this word in the generative process underlying the production of the texts. We say that the word $i$ has rank $r$ if it is the $r-th$ most frequent word. 

\subsection{Zipf's law}\label{ssec.zipf}

\begin{figure*}[!htbp]
\centering
\includegraphics[width=1\textwidth]{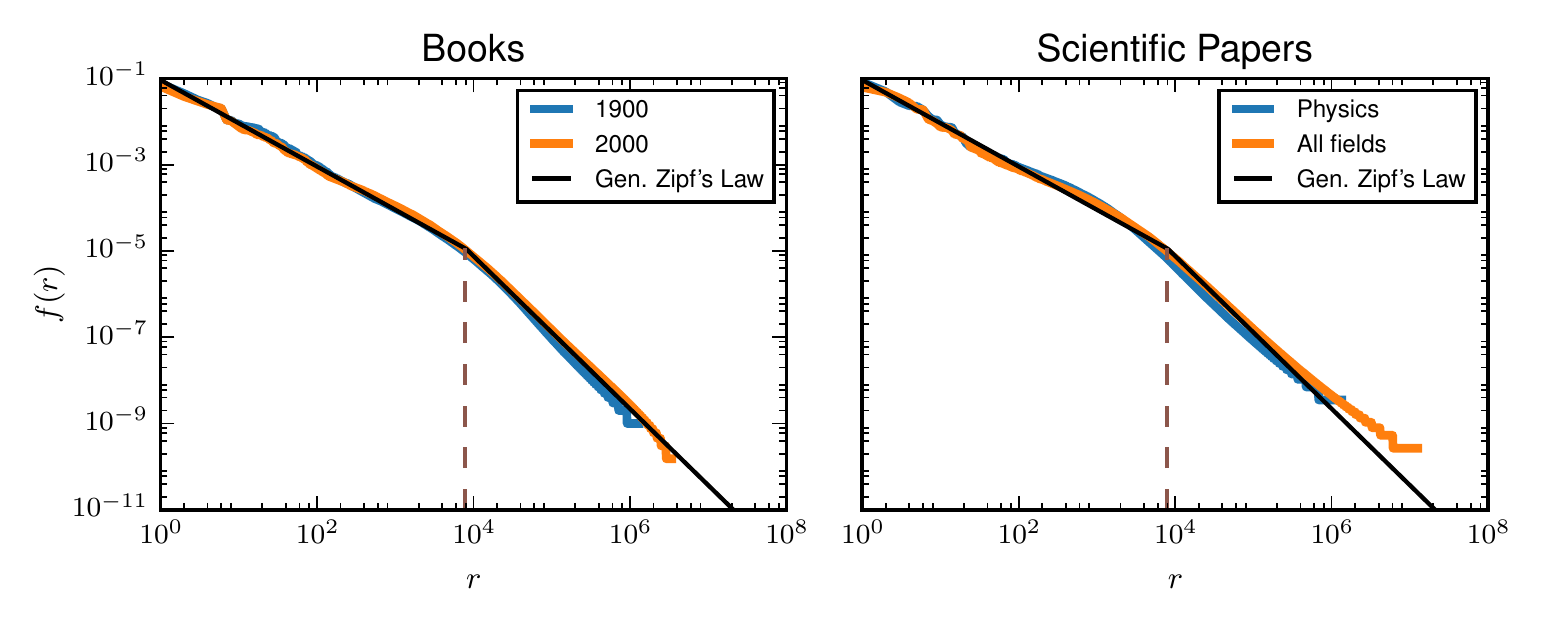}\\
\caption{Frequencies of words are distributed over a variety of scales and are well described by the modified Zipf's law. 
The (thin) black line corresponds to Eq.~(\ref{eq.modeldp}) with $b=7873$ and $\gamma=1.77$~\cite{gerlach.2013}. 
The (thick) colored lines correspond to the frequency of words obtained in different databases. (Left) Results for books published in the years $1900$ and $2000$ (see legend), as provided by the Google N-gram Database; (Right) Results for the abstract of scientific papers indexed in the Web of Science between $1991$ to $2014$ (in Physics and in all fields, see legend).}
\label{fig:distance_matrices}\label{fig.1}
\end{figure*}

Different databases show similar distributions of word frequencies, a statistical regularity also known as Zipf's law. While Zipf originally proposed the simple relationship $f(r) \propto 1/ r$, more recent analysis in large text databases suggest that the data is better described by a double power-law (dp) distribution~\cite{ferrer.2001b,petersen.2012b,gerlach.2013,williams.2014}
\begin{equation}\label{eq.modeldp}
f(r) = F^{(dp)}(r;\gamma,b)=C
\begin{cases}
r^{-1}, & r < b\\
b^{\gamma-1}r^{-\gamma} & r\geq b,
\end{cases}
\end{equation}
where $b$ and $\gamma$ are free parameters, $C=C(\gamma,b)$ is the normalization constant (which can be approximated as $C \approx 1/(G^1_{b-1}+1/(\gamma-1))$, and $G^{a}_{b} \equiv \sum_{r=1}^{b} r^{-a}$ is the $b$-th generalized Harmonic number~\cite{WebPageWolframMath}. The more common single-power-law distribution is recovered for $b\rightarrow 1$ and our results below apply in this limit as well. In plots and numerical calculations we use the distribution~(\ref{eq.modeldp}) with $b=7873, \gamma=1.77,$ and $C=0.0922$, values obtained in Ref.~\cite{gerlach.2013} for English books published in different centuries. In Fig.~\ref{fig.1} we
show that the modified Zipf's law indeed provides good account of different databases.

\subsection{Generalized Entropies}\label{ssec.halpha}

In line with the long-tradition of Information Theory, we use entropies to quantify the amount of information contained in written texts. Here we consider the generalized entropy of order $\alpha$~\cite{havrda.1967}
\begin{equation}
\label{eq.halpha}
  H_{\alpha}(\ff) = \frac{1}{1-\alpha} \left(\sum_{i=1}^M (f_i)^{\alpha} - 1 \right),
\end{equation}
where $\ff = (f_1, f_2, \ldots, f_M)$, the sum runs over all words for which $f_i \ne 0$, and $\alpha$ is a free parameter yielding a spectrum of entropies. For $\alpha=1$ we recover the Gibbs-Shannon entropy, i.e. $H_{\alpha=1} = - \sum_i f_i \log f_i$. In Physics, Eq.~(\ref{eq.halpha}) is known as Tsallis entropy~\cite{gellmann.book2004,BookByTsallis} and has been proposed as a (non-extensive) generalization of the traditional Statistical Mechanics.

\subsection{Divergence Measures}\label{ssec.dalpha}

We are particularly interested in using $H_\alpha$ to quantify the distance (or dissimilarity) between different databases.
Here we focus on the generalized Jensen-Shannon divergence~\cite{burbea.1982}
\begin{equation}
\label{eq.jsd}
 D_\alpha(\pp,\qq) = H_\alpha\left(\frac{\pp +\qq}{2}\right) - \frac{1}{2} H_\alpha(\pp) - \frac{1}{2}H_\alpha(\qq),
\end{equation}
where $\pp$ and $\qq$ are the word frequencies of the two databases and $\pp+\qq = \sum_i p_i + q_i$ is obtained summing over all symbols for which either $p_i \ne 0$ or $q_i \ne 0$. We focus on $D_\alpha$ because $\sqrt{D_q}$ can be shown to be a metric for $0 < \alpha \le 2$, i.e., it is positive $D_\alpha \ge 0$ (with $D_\alpha=0$ if and only if $\pp=\qq$), symmetric $D_\alpha(\pp,\qq) = D_\alpha(\qq,\pp)$, and $\sqrt{D_{\alpha}}$ satisfies the triangular inequality~\cite{grosse.2002,endres.2003,briet.2009}. We expect our main results to apply also to other quantities obtained from $H_\alpha(\pp,\qq), H_\alpha(\pp), $ and $H_\alpha(\qq)$,
such as the generalized Mutual Information and Kullback-Leibler divergence~\cite{cover.book2006}. The usual ($\alpha=1$, Jensen-Shannon) divergence  is a traditional method in different statistical analysis of natural language~\cite{manning.book1999}. For generalized entropies, increasing (decreasing) $\alpha$ one increases (decreases) the weight of the most frequent words allowing for different insights into the relationship between the databases~\cite{Gerlach.2016}.

\section{Effect of Zipf's law on Generalized Measures}

The goal of this paper is to investigate the consequences of known properties of word statistics to the computation of generalized entropic measures. For instance, the number of different words is virtually unbounded and therefore we should carefully consider finite-size effects and the role played by the number of observed symbols in our analysis~\cite{dodds.2011,Gerlach.2016}. More specifically, we explore the consequences of Zipf's law -- as reviewed in Sec.~\ref{ssec.zipf} -- to the computation of the information-theoretic measures based on $H_\alpha$ -- reviewed in Secs.~\ref{ssec.halpha} and~\ref{ssec.dalpha}. 
In Ref.~\cite{Gerlach.2016} we have shown that Zipf's law implies that finite-size estimators of $H_\alpha$ and $D_\alpha$ scale very slowly with database size. Here we focus on the contribution of individual words to $H_\alpha$ and $D_\alpha$, showing how different frequency ranges dominate the estimation for different values of $\alpha$.

\subsection{Entropy $H_\alpha$}

\begin{figure*}[!htbp]
\centering
\includegraphics[width=0.34\textwidth]{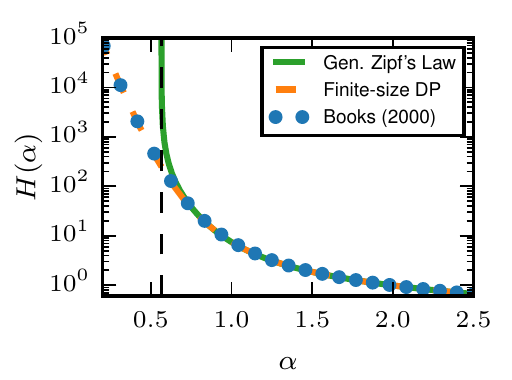}\includegraphics[width=0.34\textwidth]{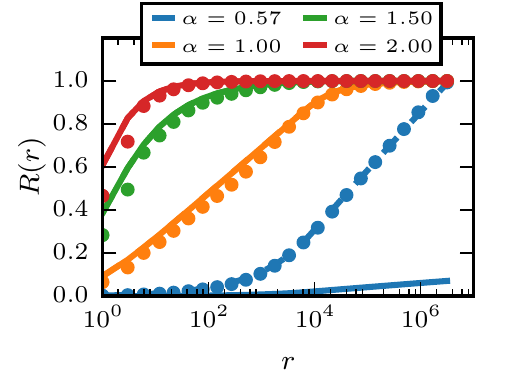}\includegraphics[width=0.34\textwidth]{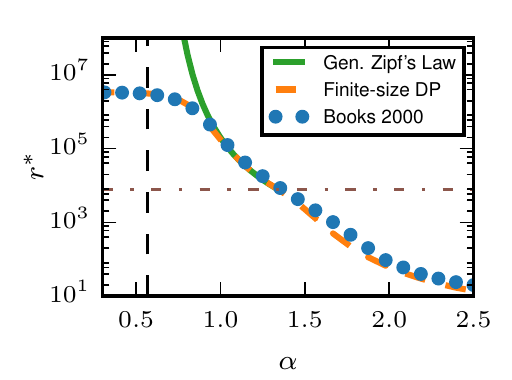}
\caption{Contribution of the $r$ most frequent words to the estimation of the generalized entropy $H_\alpha$. Symbols are the results obtained for the data (books published in the year 2000). Lines are the theoretical predictions from the double-power-law distribution~(\ref{eq.modeldp}) with the same number of words as in the data (dashed line, finite-size DP) and with infinite support (solid line, obtained analytically). (a) $H_\alpha$ as a function of $\alpha$, solid line corresponds to Eq.~(\ref{eq.hdp}); (b) Contribution of the $r$ most frequent words measured by the ratio $R_\alpha(r)$ given in Eq.~(\ref{eq.R}), solid lines correspond to Eq.~(\ref{eq.Rdp}); and (c) The rank $r^*$ for which $R_\alpha(r=r^*)=99\%$, solid line corresponds to Eqs.~(\ref{eq.rstarhss1})-(\ref{eq.rstarh}). }
\label{fig.2}
\end{figure*}

The entropy~(\ref{eq.halpha}) is uniquely defined by the frequency of the words $\ff$. From the double power-law (dp) frequency distribution, Eq.~(\ref{eq.modeldp}), we obtain
\begin{equation}\label{eq.hdp}
H_\alpha^{(dp)} \equiv \frac{1}{1-\alpha} \left (\sum_{r=1}^{\infty} (F_{dp}(r))^\alpha-1\right) = \frac{1}{1-\alpha} \left( C^\alpha(h_1+h_2)-1 \right),
\end{equation}
with
$$h_1 = \sum_{r=1}^{b-1} r^{-\alpha} \equiv G_{b-1}^{\alpha} \; \text{ (generalized Harmonic number)},$$
%
and
$$ h_2 = b^{\alpha(\gamma-1)} \sum_{r=b}^\infty r^{-\alpha \gamma} = b^{\alpha(\gamma-1)} \left( \zeta(\alpha\gamma) -G_{b-1}^{\alpha\gamma} \right) \approx \frac{b^{1-\alpha}}{\alpha \gamma -1},$$
where $\zeta(a)$ is the Riemann zeta function and the right hand side is obtained approximating the sum by the integral and is valid for $\alpha>1/\gamma$ (where $H_{\alpha}<\infty$). 
The divergence of $H_\alpha$ for $\alpha \le 1/\gamma$ appears because the sum/integral diverges for $r\rightarrow \infty$ (i.e., for a growing number of different words).
A comparison between $H_\alpha$ in real data and $H_\alpha^{(dp)}$ is shown in Fig.~\ref{fig.2}(a). The difference between the theory and the data for $\alpha \lessapprox \alpha_{c} = 1/\gamma$ is due to the finite number of symbols in the database. This is a finite-size effect that depends sensitively on the size of the database used to estimate $\ff$.

We now focus on the contribution of individual words for $H_\alpha$. To do that, we take advantage of the fact that $H_{\alpha}$ can be written as a sum over different words and consider the ratio
\begin{equation}\label{eq.R}
R(r) = \frac{\sum_{r'=1}^r (f_{r'})^\alpha}{\sum_{r'=1}^{\infty} (f_{r'})^\alpha}
\end{equation}
as a proxy for the contribution of the first $r$ terms to the computation of $H_\alpha$.  For the case of the double power-law distribution $f_r=F_{dp}(r)$, we obtain that 
\begin{equation}\label{eq.Rdp}
(h_1+h_2) R^{(dp)}(r) = \begin{cases}
\sum_{r'=1}^r r'^{-\alpha}, & = G_r^{\alpha}, \text{ for }  r < b\\
\sum_{r'=1}^{b-1} r'^{-\alpha} + b^{\alpha(\gamma-1)} \sum^r_{r'=b} r'^{-\gamma \alpha}& 
   = G_{b-1}^\alpha + b^{\alpha(\gamma-1)} ( G_r^{\alpha \gamma} - G_{b-1}^{\alpha \gamma} )  \text{ for } r\geq b,
\end{cases}
\end{equation}
For $r>b$ we can approximate the sum $\sum^r_{r'=b} r'^{-\gamma \alpha}$ by an integral and obtain
\begin{equation}\label{eq.RHalpha}
R^{(dp)}(r) \approx \frac{1}{h_1+h_2} \left( h_1+ \frac{b^{\alpha(\gamma-1)}}{\alpha \gamma -1} (b^{1-\alpha\gamma} - r^{1-\alpha\gamma})\right).
\end{equation}
In Fig.~\ref{fig.2}(b) we show the dependence of $R$ and $R^{(dp)}$ on $r$ for different values of $\alpha$. A deviation due to finite-size effects is again observed when $\alpha \rightarrow 1/\gamma$ (finite database size).

The analysis of $R$ reveals a convergence that varies dramatically with $\alpha$ (see also Refs.~\cite{dodds.2011,Gerlach.2016}), suggesting that for different $\alpha$'s different ranges in $f$ contribute to $H_\alpha$. One quantity of interest is the rank $r_{q}^*$ so that $r\le r_{q}^*$ accounts for a fraction $q$ of the effect, e.g., for $q=0.99$ we have that $R(r_q^*)=0.99$ meaning that the first $r_q^*$ terms are responsible for $99\%$ of the total $\sum_r f_i^\alpha$. For small $q$ or large $\alpha$, such that $r_q^*<b$, $r_q^*$ is obtained from the first line of Eq.~(\ref{eq.Rdp}) as the solution of 
\begin{equation}\label{eq.rstarhss1}
G_{r_q^*}^\alpha = q.
\end{equation}
For large $q$ or small $\alpha$, such that $r_q^*>b$, $r_q^*$ can be obtained explicitly from Eq.~(\ref{eq.RHalpha}) as
\begin{equation}\label{eq.rstarh}
r^{*}_q(\alpha) = \left(b^{1-\alpha \gamma} -\frac{\alpha\gamma-1}{b^{\alpha(\gamma-1)}} (q h_2-(1-q) h_1) \right)^{1/(1-\alpha\gamma)}.
\end{equation}
The estimations~(\ref{eq.rstarhss1}) and~(\ref{eq.rstarh}), which are based on the double power-law distribution~(\ref{eq.modeldp}), and the results obtained in the data are shown in Fig.~\ref{fig.2}(c). We see that for $\alpha=1$ one typically needs around 200,000 different word types in order to obtain $99\%$ of the asymptotic value of $R$. This number quickly decays with $\alpha$ so that for $\alpha=2$, the 100 most frequent words lead to the same relative contribution and therefore all other words are irrelevant in practice.

\begin{figure*}[!htbp]
\centering
\includegraphics[width=0.34\textwidth]{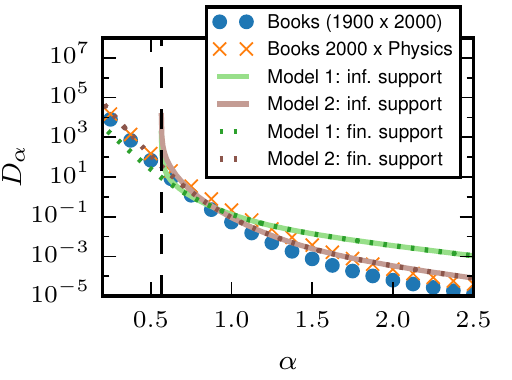}\includegraphics[width=0.34\textwidth]{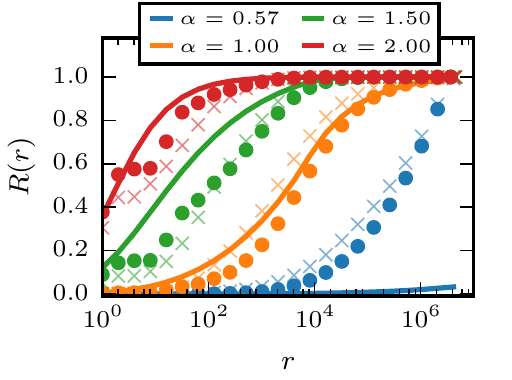}\includegraphics[width=0.34\textwidth]{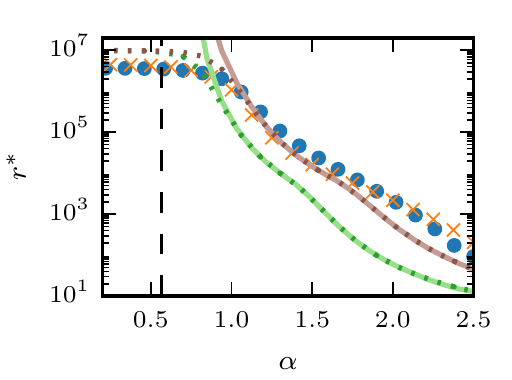}
\caption{Contribution of the $r$ most frequent words to the estimation of the generalized divergence $D_\alpha$. 
Symbols are the results obtained for the data: books published in 1900 vs. books published in 2000 (dots) and books published in 2000 vs. abstracts of Web of science papers (crosses). 
Lines are the theoretical predictions from the double-power-law distribution~(\ref{eq.modeldp}) with infinite support assuming $\Delta_i \propto f_i$, Eq.~(\ref{eq.A}) (light solid line, model 1),  and $\Delta_i \propto f_i \log f_i$, Eq.~(\ref{eq.Alog}) (dark solid line, model 2). (a) $D_\alpha$ as a function of $\alpha$; (b) Contribution of the $r$-most frequent words (ranked by the average frequency); and (c) The rank $r^*$ for which $R_\alpha(r=r^*)=99\%$. 
}
\label{fig.3}
\end{figure*}

\subsection{Divergence $D_\alpha$}

The divergence $D_\alpha$ defined in Eq.~(\ref{eq.jsd}) quantifies how dissimilar two databases are ($\pp$ and $\qq$) and the distribution of frequencies in these databases alone does not specify $D_\alpha$ . Still, we expect the general shape of Zipf's law in Eq.~(\ref{eq.modeldp}) to affect the statistical properties of $D_\alpha$. Here we explore this connection by following steps similar to those performed in the previous section for $H_\alpha$. To do this, it is convenient to introduce the relative coordinates $f_i, \Delta_i$, where  $f_i = (p_i + q_i)/2$ and $\Delta_i = \left|p_i - q_i\right|/2$, such that:
\begin{equation}\label{eq.Ddelta}
D_\alpha(\pp,\qq) = D_\alpha(\ff, \dd) =  \sum_i \frac{1}{1-\alpha} \left( (f_i)^\alpha -\frac{1}{2}(f_i +\Delta_i)^\alpha - \frac{1}{2} (f_i - \Delta_i)^\alpha\right) \equiv \sum_r D_\alpha(r).
\end{equation}
This equation emphasizes that $D_\alpha$ is computed as a sum over a contribution $D_\alpha(r)$ of different words ranked by $r$. We order the words according to the rank $r$ of the word in $\ff$, i.e., if a word has rank $r'$ it means that there are exactly $r'-1$ other words for which the average frequency $f_r=(p_r+q_r)/2 >f_{r'} =(p_{r'}+q_{r'})/2$.  

The relative contribution $\mathbbm{R}(r)$ of the top $r$ words to $D_\alpha$ is given by
\begin{equation}\label{eq.R2}
\mathbbm{R}(r) = \frac{\sum_{r'=1}^{r}D_\alpha (r')}{D_\alpha}  =  \frac{\sum_{r'=1}^{r} \left( (f_{r'})^\alpha -\frac{1}{2}(f_{r'} +\Delta_{r'})^\alpha - \frac{1}{2} (f_{r'} - \Delta_{r'})^\alpha\right)}{\sum_{r'=1}^{\infty} \left( (f_{r'})^\alpha -\frac{1}{2}(f_{r'} + \Delta_{r'})^\alpha - \frac{1}{2} (f_{r'} - \Delta_{r'})^\alpha\right)}, 
\end{equation}
which is analogous to Eq.~(\ref{eq.R}) but in this case $D_\alpha(r)$ is not necessarily monotonically decaying with $r$. We finally define $\mathbbm{r}^*_q$ as the rank at which a fraction $q$ of the total $D_\alpha$ is achieved, i.e. $\mathbbm{R}(\mathbbm{r}^*_q) = q$. 

Figure~\ref{fig.3} shows our analysis of the divergence ($D_\alpha$, $\mathbbm{R}(r)$, and $\mathbbm{r}^*_q$) for two pairs of databases (Books2000--Books1900 and Books2000--Physics, see caption of Fig.~\ref{fig.1} for details on the data). The left panel shows that the divergence $D_\alpha$ for Books2000--Physics is systematically larger than for Books2000--Books1900 suggesting that stylistic and topical differences between books and scientific papers are more significant than historical changes in the language throughout the 20-th century. 
The most striking feature of Fig.~\ref{fig.3} is the similarity between the results obtained with different data (e.g., the variation across the databases is much smaller than the variation across $\alpha$ or $r$). 
Furthermore, the general behavior observed for $D_\alpha$ resembles the results shown in Fig.~\ref{fig.2} for $H_\alpha$, which were analytically computed from the word-frequency distribution~(\ref{eq.modeldp}). 
The $D_\alpha$-observation, however, depends not only on the word frequencies~$f_i$ but also on the variation $\Delta_i$ across databases. 
Next we consider two very simplistic models for $\Delta_i$ in order to understand these observations.

\paragraph{Constant relative fluctuation.} A simple assumption is that 
the relative fluctuations across databases are the same for each word independent of its frequency, in which case $\dd$ is proportional to the average frequencies $\ff$ and thus
\begin{equation}\label{eq.A}
\frac{\Delta_i}{f_i} = A.
\end{equation}
In this case we obtain from~(\ref{eq.Ddelta}) that
\begin{eqnarray}\label{eq.DH}
D_\alpha &=  \left(1-\frac{1}{2}(1-A)^\alpha  -\frac{1}{2}(1+A)^\alpha \right)\frac{1}{1-\alpha} \sum_r (f_r)^\alpha  \\
        &= \left(1-\frac{1}{2}(1-A)^\alpha  -\frac{1}{2}(1+A)^\alpha \right) \left(H_\alpha(\ff)+\frac{1}{1-\alpha}\right)\\
        &\approx \frac{\alpha(1-\alpha)}{2} A^2 \left(H_\alpha(\ff)+\frac{1}{1-\alpha}\right),
\end{eqnarray} 
where the approximation is valid for $A \ll 1$.
Now we notice that $\ff$ is the word frequency distribution of the combined database and that therefore it should also be well approximated by the generalized Zipf's law~(\ref{eq.modeldp}). Even if this model is too simplistic to account for the observed $D_\alpha$ (see dotted line in the left panel of Fig.~\ref{fig.3}), it shows how the statistical properties of $D_\alpha$ and of $H_\alpha$ can be connected to each other.  

\begin{figure*}[!htbp]
\centering
\includegraphics[width=1\textwidth]{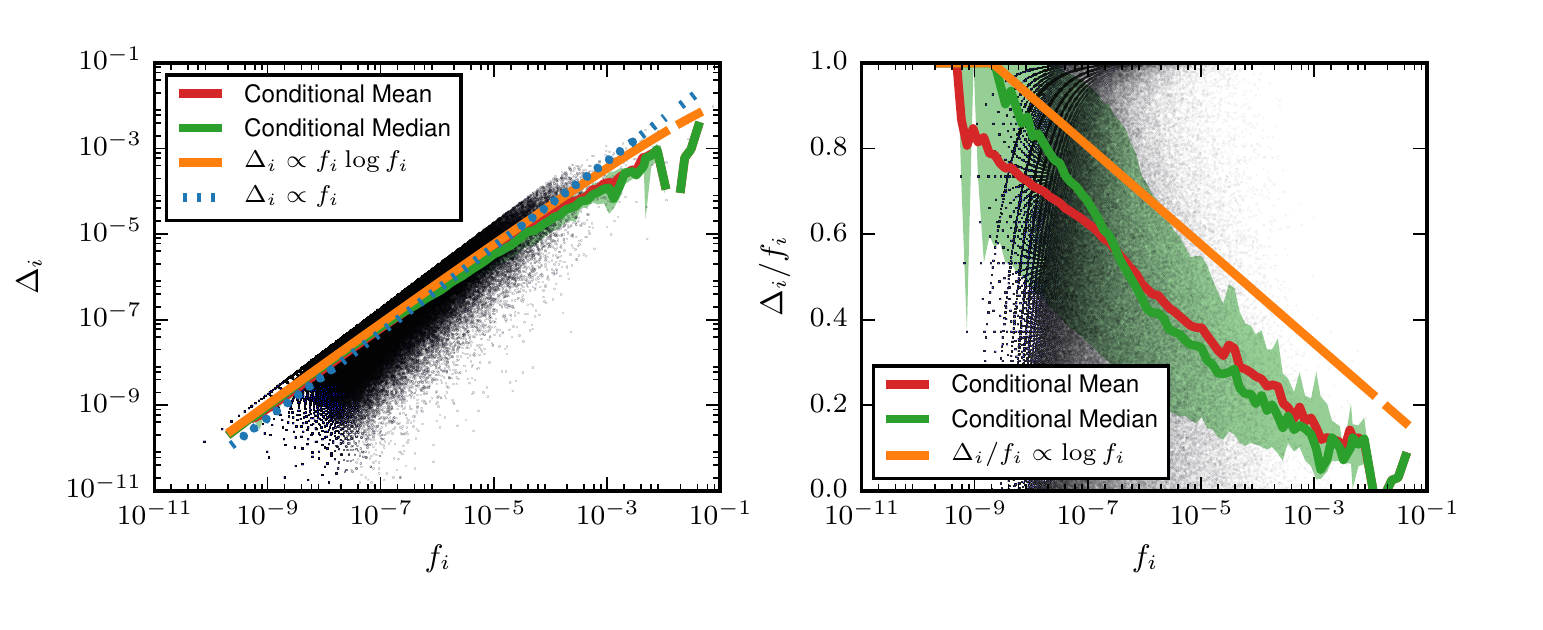}\\
\includegraphics[width=1\textwidth]{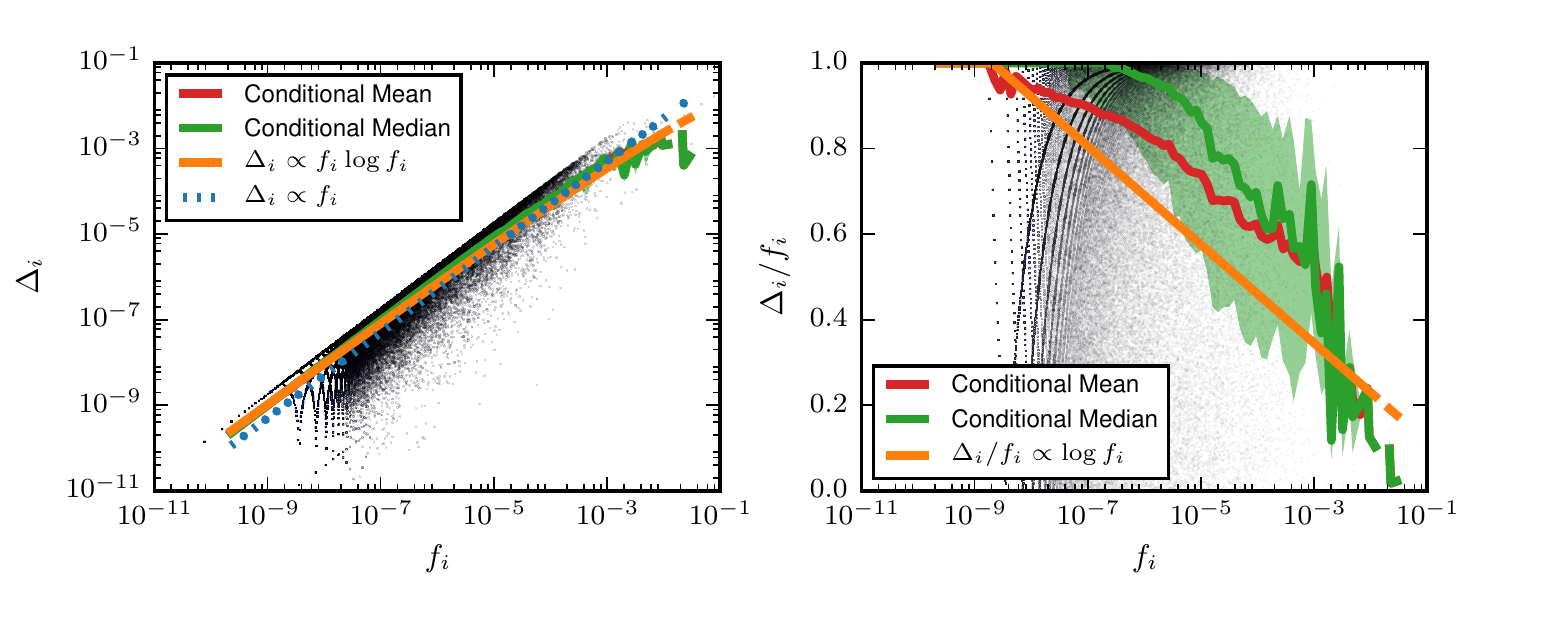}
\caption{Relation between relative $\Delta_i = \left|p_i - q_i\right|/2 $and average $f_i = (p_i + q_i)/2$ frequency.
Mean and median (conditioned on window in $f_i$) are shown for divergences between books published in the year 1900 and 2000 (top panels) and books published in 2000 and abstracts from WoS (bottom panels). Shaded region correspond to $25$- and $75$-percentile. Approximations for the conditional mean are given by $\Delta_i/f_i = 0.5$ (dotted line) and $\Delta_i/f_i = -0.05 \log f_i$ (dashed line).
}
\label{fig.4}
\end{figure*}

\paragraph{Log-corrected fluctuations.} In order to get some insights on the reason for the failure of the previous model, we look at the empirical relative fluctuation $\frac{\Delta_i}{f_i}$ for the two pairs of databases described above. 
The results in Fig.~\ref{fig.4} show two features: an expected large fluctuation around different words and a surprising decay of relative fluctuation with $f_i$. The roughly linear decay in the semi-logarithmic plot suggests that an improvement of Eq.~(\ref{eq.A}) is obtained including a logarithmic correction as $\Delta_i/f_i \propto \log f_i$. 
Since $\Delta_i$ is bounded from above by $f_i$ (i.e. $\Delta_i \leq f_i$) we introduce a lower cutoff frequency in our log-corrected model
%
\begin{equation}\label{eq.Alog}
\frac{\Delta_i}{f_i} = 
\begin{cases}
a \log f_i/f_{\max} &, f>f_{\max}e^{1/a}\\
1 &, f\leq f_{\max}e^{1/a}
\end{cases},
\end{equation}
where we empirically find that $f_{\max}=1$ and $a=-0.05$ capture the main qualitative behaviour shown in Fig.~\ref{fig.4}.
The log-corrected model, obtained combining Eq.~(\ref{eq.Alog}) with the generalized Zipf's law~(\ref{eq.modeldp}), provides a much better account of the results in the three panels of Fig.~\ref{fig.3}.
This shows that the weak dependence of the relative fluctuations on the frequency is crucial in order to understand the results in Fig.~\ref{fig.3}.

\section{Implication of our results}

\subsection{Keywords in Physics}

Our results shows that the Zipf's law is responsible for the general statistical properties of both $H_\alpha$ and $D_\alpha$. One consequence of this result is that the contribution of (a set of) particular words is also pre-determined by Zipf's law and depends largely on the range of frequencies of the words. Consider the problem of comparing the divergence between the corpus of scientific papers in Physics to a general corpus of books written in English. One of the effects one may want to capture when computing $D_\alpha$ is the over-representation of physics-related words in the database of Physics articles, i.e., the fact that $p_i > q_i$ for words $i$ related to Physics. We denote this set of words as physics keywords. This is not the only effect contributing to the divergence $D_\alpha$ between the texts, e.g., stylistic effects affecting the most frequent words (so-called stopwords) may also be relevant. Here we wish to quantify the effect of Physics keywords to $D_\alpha$ in comparison to a set of stopwords.

\begin{figure*}[!htbp]
\centering
\includegraphics[width=1\textwidth]{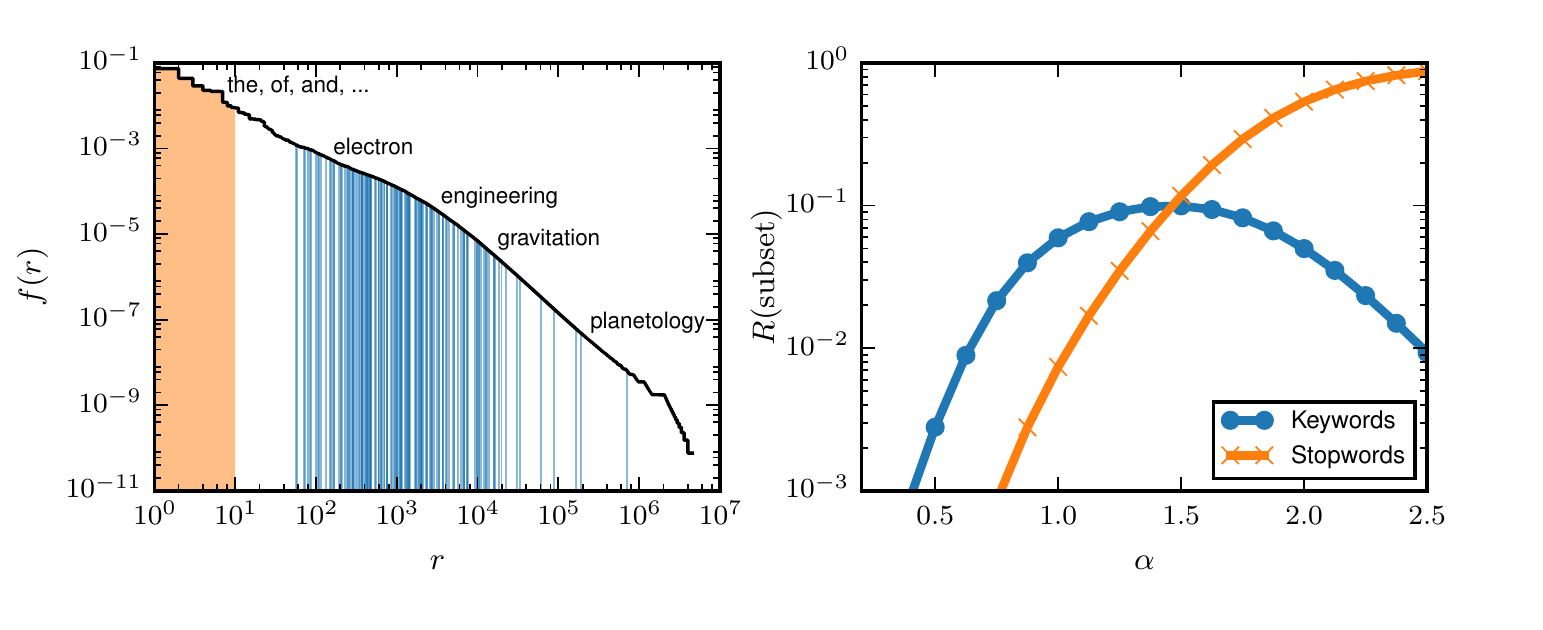}
\caption{Contribution of subsets of words to the divergence $D_\alpha$. Results are shown for a list of $318$ physics keywords (see text) and a list of the $10$ most frequent stopwords (the, of, and, in, to, a, is, for, that, with).
(Left) Position of keywords and stopwords in the rank-frequency distribution. 
(Right) Fraction of the generalized divergence $D_{\alpha}$ from words belonging to the list of keywords and the list of stopwords as a function of $\alpha$.
}
\label{fig.pacs}\label{fig.5}
\end{figure*}

The key insight that connects this problem to our results is that Physics keywords are typically distributed in a specific range of frequencies. For instance, we compiled a list of $318$ Physics keywords from all words appearing in the PACS system (removing a list of common stop words). As illustrated in the Fig.~\ref{fig.5}(left panel) the words range from {\it electron} -- with rank $r_i \approx 100$ and frequency of one every thousand words $f_i \approx 10^{-3}$ --  to {\it gravitation} -- with rank $r_i \approx 2000$ and frequency of one every hundred thousand words $f_i \approx 10^{-5}$. Most Physics keywords lie in between these two frequencies. By increasing $\alpha$ from $\alpha = \alpha_c =1/\gamma \approx 0.56$ one moves from a configuration in which $D_\alpha$ and $H_\alpha$ are dominated by the least frequent words to a configuration in which $D_\alpha$ and $H_\alpha$ are determined mostly by the most frequent stopwords (e.g., for $\alpha > 2$). Indeed, the results in Fig.~\ref{fig.5}(right panel) confirm that the contribution of the Physics keywords has a maximum around $\alpha\approx 1.4$. At the maximum, these $318$ keywords contribute with more than $10\%$ of the total value of $D_\alpha$. This value is comparable to the contribution of the $10$ most frequent words (stopwords) at the same value of $\alpha$. The contribution of the stopwords quickly increases with $\alpha$ and completely dominates $D_\alpha$ for $\alpha \gtrapprox 2.0$

\subsection{How large does my database have to be?}

When computing $H_\alpha$ and $D_\alpha$ one usually aims at characterizing the properties of the source (stochastic process) underlying the data. Stationarity and ergodicity of this process imply that computed values should converge for increasing database size. In practice, we are not interested in results which depend mainly on the size of the database, and that change dramatically with the amount of available data. Below we show how our results allow for an estimation of the database size required to provide a reliable estimation of $D_\alpha$.

\begin{figure*}[!htbp]
\centering
\includegraphics[width=0.5\textwidth]{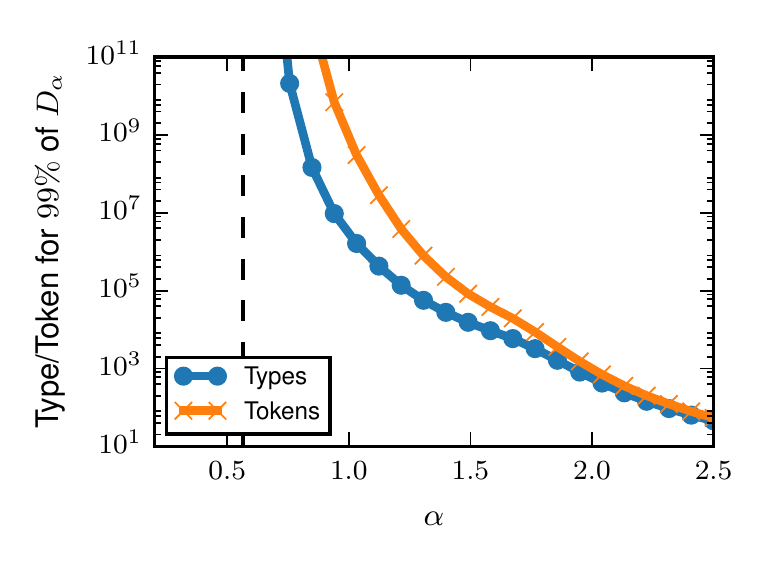}
\caption{
Database size necessary to observe $99\%$ of $D_{\alpha}$. The curve for the number of different words (types) $M$ was computed from $\mathbbm{r}$ as in Fig.~\ref{fig.3}. The relationship $M \sim N^{1/\gamma}$ to the size of the database $N$ (number of tokens) was obtained from a Poisson null model assuming a double power-law Zipfian distribution, as in Ref.~\cite{gerlach.2014}. For comparison, the typical book size in Project Gutenberg is $N\approx 10^5$, implying that  $D_\alpha$ between two books can typically be computed only for $\alpha>1.5$. }
\label{fig.heaps}\label{fig.6}
\end{figure*}

The most important effect of changing the database size is to increase the number of different words found in the databases. This simple observation, the cornerstone of our analysis, has two ramifications. First, it implies that a necessary condition for a robust estimation of $D_\alpha$ is that $M>\mathbbm{r}_{q\rightarrow 1_-}^*$, i.e. the number of observed different words $M$ needs to be larger than the number of ranks $r$ needed to estimate a fraction $q\lessapprox 1$ of $D_\alpha$. Second, a connection to the size of the database~$N$ (measured in number of word tokens) is possible through Heaps' law, 
which states that the number of different words grows sublinear with the total number of words, ~$M \sim N^{1/\gamma}$ ~\cite{herdan.book1960,heaps.book1978}. 
In Fig.~\ref{fig.6} we present the result of this analysis, in which $\mathbbm{r}^*_{q=0.99}$ was obtained from the double-power-law distribution with log-corrected fluctuations (as in Fig.~\ref{fig.3}) and the Heaps' law relationship derived in Ref.~\cite{gerlach.2014}.

\section{Discussion and Conclusions}

The main message of this paper is that the characteristic shape of word-frequency distributions ($f_r$ following Zipf's law) plays a dominant role in the properties of information-theoretic measures computed in texts at the level of words. While there is a one-to-one relationship between $f_r$ and entropies $H_\alpha$ -- given in Eq.~(\ref{eq.hdp}) -- here we showed that a close connection exists also between $f_r$ and measures intended to compare databases such as $D_\alpha$, a result that presumably extends also to other measures such as the Mutual Information and Kullback-Leibler divergence. The influence of $f_r$ occurs not only in the convergence of finite-size estimators, as reported previously in Refs.~\cite{dodds.2011,Gerlach.2016}, it affects the value of $D_\alpha$ and the weight of the contributions of words in different frequency ranges. This connection relies not only on the universality of $f_r$ but also on our empirical finding that, for different pairs of databases, the relative fluctuations decay with the logarithm of the frequency, see Eq.~(\ref{eq.Alog}) and Fig.~\ref{fig.4}.

The finding that Zipf's law directly controls the expected weights of contribution of different words provides a further motivation for our choice of using generalized entropies $H_\alpha$. The variation of the free parameter $\alpha$ effectively tunes the range of frequency of the words that contribute to $H_\alpha$ and $D_\alpha$: for large $\alpha$ (e.g., $\alpha=2$) only the most frequent words contribute, while for $\alpha<1$ the results are dominated by the least frequent words. From an example based on $318$ keywords in Physics, we obtain that these words contribute with $6\%$ of $D_{\alpha=1}$, $10\%$ of $D_{\alpha=1.4}$, but only $5\%$ of $D_{\alpha=2}$.  Words in different frequency ranges have different semantic and syntactic properties so that the variation of $\alpha$ can characterize also different types of divergencies between the databases.

As $\alpha$ is reduced and approaches (from above) the critical value $\alpha=1/\gamma$, where $\gamma$ is the exponent of Zipf's law defined in Eq.~(\ref{eq.modeldp}), the convergence of $H_\alpha$ and $D_\alpha$ becomes extremely slow and increasingly large text sizes are needed for a robust estimation (see Fig.~\ref{fig.6}). For instance, for the usual Jensen-Shannon divergence  $D_{\alpha=1}$ we estimate that databases of size  $\approx 10^{8}$ tokens ($\approx 200$ book or $\approx 10^{6}$  word types) is needed  while for $\alpha=0.6$ the size grows dramatically to the unrealistic number of $ \approx10^{20}$ tokens ($\approx 2 10^{14}$ books or $\approx 10^{16}$ word types). For $\alpha<1/\gamma \approx 0.56$ there is no convergence and therefore these quantities are not properly defined. This is one of the most dramatic consequences of Zipf's law and reflects the effectively unbounded number of different symbols (vocabulary) in which $H_\alpha$ is computed.

\section*{Acknowledgments} L.D. was funded by CAPES (Brazil). E.G.A. and M. G. are grateful to F. Font-Clos for helpful discussions on the subject of this manuscript.

\section*{References}
\providecommand{\newblock}{}

\end{document}